\documentclass[sigconf, 10pt]{acmart}    
\AtBeginDocument{%
  }

\setcopyright{acmlicensed}
\copyrightyear{2024}
\acmYear{2024}
\acmDOI{XXXXXXX.XXXXXXX}

\acmConference[PICASSO'24]{MobiCom'24}{November 18,
  2024}{Washington, D.C., USA}
\acmISBN{978-1-4503-XXXX-X/18/06}
\usepackage{dsfont}

\begin{document}

\title[Wiener-Denoiser Network for Misaligned OAC]{Misaligned Over-The-Air Computation of Multi-Sensor Data with Wiener-Denoiser Network}

\author{Mingjun Du}
\email{dumj23@mails.tsinghua.edu.cn}
\affiliation{%
     \institution{Tsinghua University}
  \city{Shenzhen}
  \country{China}
}

\author{Sihui Zheng}
\email{zhengsh21@mails.tsinghua.edu.cn}
\affiliation{%
      \institution{Tsinghua University}
  \city{Shenzhen}
  \country{China}
}

\author{Xiao-Ping Zhang}
\authornote{Corresponding author}
\email{xpzhang@ieee.org}
\affiliation{%
\institution{Tsinghua University}
  \city{Shenzhen}
  \country{China}
}

\author{Yuhan Dong}
\email{dongyuhan@sz.tsinghua.edu.cn}
\affiliation{%
      \institution{Tsinghua University}
  \city{Shenzhen}
  \country{China}
}

\renewcommand{\shortauthors}{Mingjun Du et al.}

\begin{abstract}
    In data driven deep learning, distributed sensing and joint computing bring heavy load for computing and communication. To face the challenge, over-the-air computation (OAC) has been proposed for multi-sensor data aggregation, which enables the server to receive a desired function of massive sensing data during communication. However, the strict synchronization and accurate channel estimation constraints in OAC are hard to be satisfied in practice, leading to time and channel-gain misalignment. The paper formulates the misalignment problem as a non-blind image deblurring problem. At the receiver side, we first use the Wiener filter to deblur, followed by a U-Net network designed for further denoising. Our method is capable to exploit the inherent correlations in the signal data via learning, thus outperforms traditional methods in term of accuracy. Our code is available at \href{https://github.com/auto-Dog/MOAC\_deep}{https://github.com/auto-Dog/MOAC\_deep}.
 
\end{abstract}

%
%




\maketitle

\section{Introduction}

 In the era of big data, data-driven artificial intelligence models have made significant achievements in industry and life. High-quality and large-scale data from the world plays a key role for improving the performance of models\cite{fan2024scaling}. However, we are faced with the challenges of massive data transmission and computation, e.g. a big data system for environmental sensing should transmit and collect a large amount of data while minimizing data latency, a requirement that traditional systems struggle to meet\cite{Kim2019_sensorbigdata}. Therefore, it is necessary to find methods to reduce computational load and accelerate communication speed. 

 Sensor data aggregation is a scenario of multiple sensors accessing a central server base station (BS) via wireless communication. Traditionally, this is done by allocating communication resources to each sensor, e.g. TDMA, CDMA, FDMA, and then computing at the BS. But OAC can complete computing tasks while enabling communication of sensors, thereby meeting the requirements above\cite{abari2016over,zhu_over--air_2021,zhu_mimo_2019}. 
 OAC is a form of parallel communication, widely used in distributed sensing, distributed consensus, and distributed edge learning. Through specific pre-processing $\phi(\cdot)$ functions at the sensors and post-processing $\psi(\cdot)$ functions at the BS, see Eqn. \ref{equ:OACNom}, OAC can perform a series of function computations on $x_m$, including summation, maximum and minimum values, arithmetic mean, Euclidean norm and so on\cite{csahin2023survey}. These computations are carried out in the wireless channel, and the BS only needs to receive the computation results $\boldsymbol{y}$. 
 \begin{equation}
     \boldsymbol{}{y}= \psi(\sum_{m=1}^M \phi_m(\boldsymbol{x_m}))
     \label{equ:OACNom}
 \end{equation}

 However, OAC faces many challenges in practice. One of them is the misalignment in OAC. OAC often exhibits temporal misalignment and channel-gain misalignment effects, making it difficult for the BS to recover function computation results accurately\cite{shao_federated_2022,razavikia_blind_2022}. Existing approaches like maximum likelihood (ML) estimation, has high computation complexity and is very sensitive to noise, causing large mean square error (MSE)\cite{park_bayesian_2021,shao_bayesian_2021,shao_federated_2022}. On the other hand, these methods did not consider temporal and inter-sensors correlation of data, which can become prior knowledge for estimation. 

 To address misalignment OAC under noise, we proposed a Wiener-denoiser network. We innovatively formulate the signal estimation process as a non-blind image deblurring process, which enhances the estimation efficiency and introduces more strategies\cite{dong_deep_2021,zhang_CNN_learning_2017}. Our method employs Wiener filter for initial estimation, then employs deep learning method to fully exploit signal prior information and reduce the impact of noise on estimation. Experimental results show that our method can accurately estimate OAC results in different noise and misalignment environments, promise accuracy and efficiency. Additionally, our network has zero-shot ability, which can be applied to various datasets.

\section{Problem Formulation}
\subsection{System Model}
 A misaligned OAC channel includes $M$ sensors and a BS. In the simplest scenario, 
 we concern arithmetic-sum of data packets from different sensors. In a data packet, device $m$ sends $L$ symbols $s_m[l],l=1,2,...L$, where the symbol can be any real positive value, and each symbol lasts for a symbol period $T$. The symbols are modulated to signal with analog modulation, and sensor $m$'s signal is represented by $x_m(t)$.

Due to difficulties in accurately synchronizing the clocks at the sensors, the signals are overlapped and with time misalignment $\tau_m<T$. Also, a perfect channel coefficient compensation for sensors is hard, which causes residual channel-gain misalignment $h_m$ in the overlapped signals, usually presenting as a phase deviation. The misaligned signals travel through a flat and slow fading multiple access channel (MAC) and are aggregated over the air, given by Eqn. \ref{equ:MAC model}. 
 \begin{equation}
     y(t)=\sum_{m=1}^M h_{m}x_{m}(t-\tau_{m})+n(t)
     \label{equ:MAC model}
 \end{equation}
 where
 \begin{equation}
     x_{m}(t)=\sum_{l=1}^L s_{m}[l]p(t-lT)
 \end{equation}
Here $y(t)$ is the aggregated signal observed at BS. The time delays $\{\tau_m\}$ and channel-gain coefficients $\{h_m\}$ of each misaligned signal sequence can be known by the BS, since BS is able to resolve the predefined pilots of each client respectively then perform channel estimation\cite{jing_federated_2022}. $n(t)$ is complex Gaussian noise, $p(t-lT)$ is a pulse function to modulate discrete symbols $s_{m}[l]$ into continues signal. 
 
In the OAC system, an ideal observation from BS should be aligned sum of $s_m[l]$, that is: 
\begin{equation}
   s_+[l] = \sum_{m=1}^M  s_m[l], l=1,2,...,L
\end{equation}
In misaligned OAC, the BS is expected to estimate $s_+[l]$ with known $y(t)$, $\{\tau_m\}$, $\{h_m\}$ and noise statistics. The objective is to minimize MSE with estimation algorithm. Here we use bold letters to donate $L$ length vectors, and $\boldsymbol{\hat{s}_{+}} $ is the estimation of $\boldsymbol{s_{+}}$:
\begin{equation}
{\rm MSE}(\boldsymbol{\hat{s}_{+}},\boldsymbol{s_{+}}) = \frac{1}{L} \sum_{l=1}^L \left| \hat{s}_{+}[l] - s_+[l] \right|^2
\end{equation}

\begin{figure}[htbp]
  \centering
  \includegraphics[width=\linewidth]{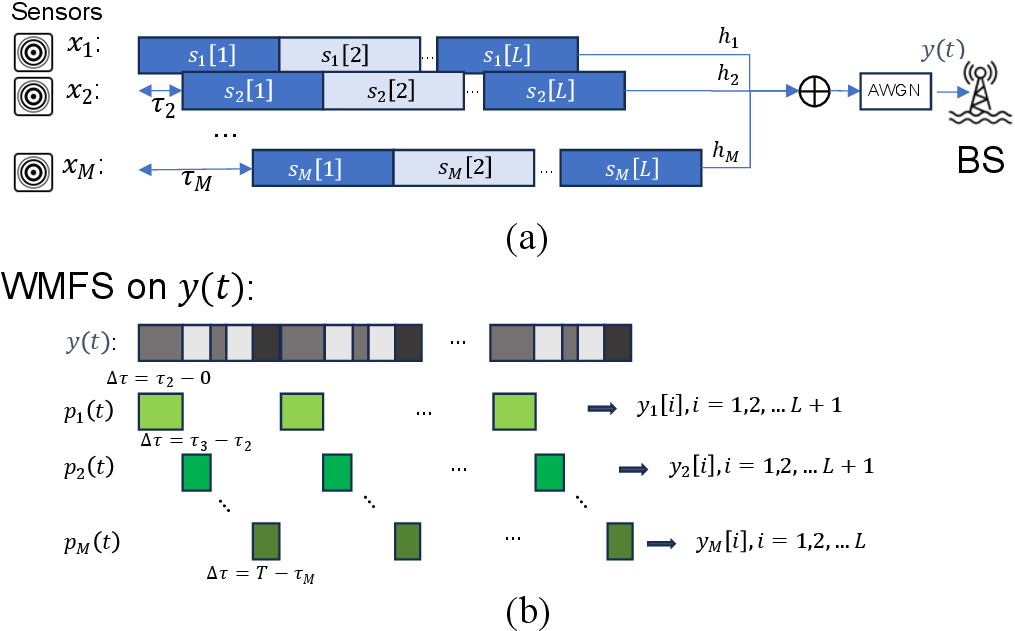}
 \caption{(a) A misaligned OAC system model. $y(t)$ is the overlapped sum of signal of $s_m[l]$. (b) WMFS process\cite{shao_federated_2022}, different colors on $y(t)$ indicate that overlap symbols are different on time slots. $p_1(t)$,$p_2(t)$... are matched filters, $y_1[i],y_2[i]...$ are samples. See Section 2.2 for details. }
  \label{fig:sys}
\end{figure}

\subsection{Estimation Problem after Sampling}
The observed signal $y(t)$ at BS contains all the combinations of overlapped symbols at different times slots within $T$, leaving the possibility of estimate all symbols in $\boldsymbol{s}$ after sampling, upon which we can further calculate the objective $\boldsymbol{\hat{s}_{+}}$. 
 
A whitened matched filter and sampling (WMFS) process for this problem was proposed by Shao et al\cite{shao_federated_2022}. Specifically, with given $y(t)$ and $\{\tau_m\}$, $M$ matched filters $p_m(t)$ are designed to filter signal, where $\Delta\tau_{m,m+1} = \tau_{m+1}-\tau_m$. 
\begin{equation}
    p_m(t) = \frac{1}{2} [sgn(t+T) - sgn(t+T-\Delta \tau_{m,m+1})] 
    \label{equ:mf}
\end{equation}
We filter $y_m(t)=y(t)*p_m(t)$ and sample at $(i-1)T + \tau_{m+1} , i =1, 2,...,L+1$, where * stands for convolution. The sampled result $y_m[i]$ is 

\begin{equation}
    \begin{aligned}
y_{m}[i]& =y_m(t=(i-1)T+\tau_{m+1}) \\
&=\frac1{\Delta\tau_{m,m+1}}\int_{(i-1)T+\tau_m}^{(i-1)T+\tau_{m+1}}\sum_{k=1}^Mh_ks_k[i-\mathds{1}_{k>m}] +n(\zeta) d\zeta\\
&\triangleq\sum_{k=1}^Mh_ks_k[i-\mathds{1}_{k>m}]+{n}_m[i]\\
\end{aligned}
\label{equ:ymi}
\end{equation}

See Fig. \ref{fig:sys} for detailed visualization. ${n}_m[i]$ is whitened and independent noise. $\mathds{1}_{k>m}$ is an indicator function, it equals to 1 when $k>m$ otherwise 0.


Eqn. \ref{equ:ymi} can be written as matrix format in Eqn. \ref{equ:sample_equ}. Now the problem is simplified into a linear estimation problem. 
\begin{equation}
   \boldsymbol{ y}=\boldsymbol{Ds}+\boldsymbol{n}
\label {equ:sample_equ}
\end{equation}

where $\boldsymbol{s}$ is all transmitted symbols in vector form,

$\boldsymbol{s} =  [s_1[1],...,s_M[1],$ $s_1[2],...s_M[2],...,s_1[M],...,s_M[M]]^{\mathsf{T}}$ 

with Gaussian noise $n$, 

$\boldsymbol{n} = 
[n_1[1],...,n_M[1],n_1[2],...,n_M[2],...,n_1[L], ...,n_M[L],$ $n_1[L+1],...,n_{M-1}[L+1]]^{\mathsf{T}}$, 

to get $M(L+1)-1$ samples $\boldsymbol{y}$

$\boldsymbol{y} = [y_1[1],...,y_M[1],y_1[2],...,y_M[2],...,y_M[1],...,y_M[L],$ $y_1[L+1],...,y_{M-1}[L+1]]^{\mathsf{T}}$. See Eqn. \ref{equ:sample_equ_D} for detail form of $\boldsymbol{D}$. 

\begin{scriptsize}
\begin{equation}
\begin{array}{ccc}
  \boldsymbol{D}=   &   \begin{bmatrix} 
    h_1&&&&&&&\\
    h_1&h_2&&&&&&\\
    ...&h_2&...&&&&&\\
    h_1&...&...&h_M&&&&\\
    &h_2&&h_M&h_1&&&&\\
    &&...&...&h_1&h_2&&&\\
    &&&h_M&...&h_2&...&&\\
    &&&&h_1&...&...&h_M\\
    &&&&&h_2&&h_M&...\\
    &&&&&&...&...&...\\
    &&&&&&&h_M&
    \end{bmatrix} _{M(L+1)-1\times ML}
\end{array}
\label {equ:sample_equ_D}
\end{equation}
\end{scriptsize}


    

\section{Existing Solutions}
\paragraph{Aligned Estimation}
The last sample at each symbol period $T$ is aligned in time domain, which means $y_M[i]=\sum_{m=1}^M h_m s_m[i]+n_m[i], i\in [1,L]$. Therefore the estimator can be written as $\hat{s}_+[i]=y_M[i]$\cite{align_UAV,shao_federated_2022}. Nevertheless, this only works when the channel gain misalignments are negligible ($h_M \rightarrow 1$), which limits its generality.
 
 \paragraph{ML estimation}
An ML estimator is given based on Eqn. (\ref{equ:sample_equ}). Since it can be seen as observation function,  a weighted least square (WLS) estimator is employed to get symbols $\boldsymbol{\hat{s}}$ \cite{ss_book}  \cite{shao_bayesian_2021}. Then, we add symbols from same period but different sensors to get $\hat{s}_+$. 
However, the method is computational demanding and sensitive to noise, that error can propagate and accumulate when computing. 


\paragraph{LMMSE estimation}
To relieve noise sensitive property of ML estimator, a linear minimum mean square error (LMMSE) method was proposed\cite{shao_bayesian_2021,liu_misaligned_2023}.
The goal is to minimize MSE with known first and second order statistics of original signal as prior information. But the statistics should be transmitted separately before each data packet, which causes additional communication. Also it computes the inverse of matrix as large as $\boldsymbol{D}$, causing computational burden. 



\section{A Wiener-Denoiser Estimator for Misaligned Samples}

In this chapter, we model the estimation problem as a non-blind image deblurring problem. Then we design our estimator which utilizes data's temporal and inter-sensors correlation as prior knowledge.

\begin{figure*}[htbp]
  \centering
  \includegraphics[width=0.8\linewidth]{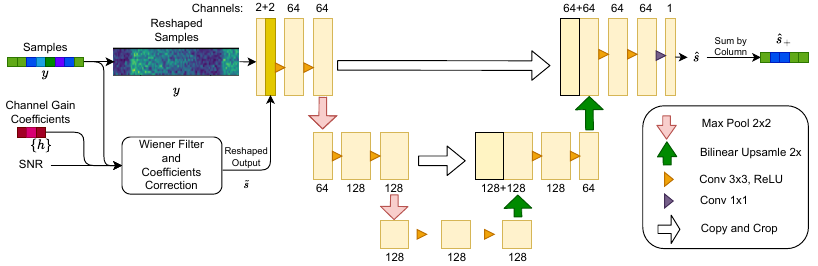}
  \caption{Framework of the proposed method. Input $\boldsymbol{y}$ passes through Wiener Filter for primary deconvolution, obtaining initial estimation $\boldsymbol{\widetilde{s}}$. The reshaped $\boldsymbol{y}$ and $\boldsymbol{\widetilde{s}}$ are concatenated and then sent into the U-Net as a four channel image for denoising. Yellow boxes are feature maps. The output is desired OAC result estimation $\boldsymbol{\hat{s}_+}$}
  \label{fig:net}
\end{figure*}

\subsection{Reformulate the Problem}
To improve calculation efficiency and avoid noise error accumulation in estimation, we exploit the property of circular-like matrix $\boldsymbol{D}$ in Eqn. \ref{equ:sample_equ_D}. If all $h_m$ are separated as column scaling factor, the sample procedure becomes $\boldsymbol{y = Ds+n }$ $\boldsymbol{ = K\Lambda_H s+n}$, where $\boldsymbol{K}$ is a matrix with same structure as $\boldsymbol{D}$ but all the non-zero elements are 1, $\boldsymbol{\Lambda_H}$ is a diagonal matrix $\Lambda_H = diag(h_1, h_2, ...h_M,h_1,...,h_M )$, shaped $ML$ $\times$ $ML$.

Furthermore, $\boldsymbol{K}$ reveals a matrix form of linear convolution, the kernel $\boldsymbol{k}$ is an all-one sequence with finite length M. We reformulate the sample procedure as follow:
\begin{equation}
   \boldsymbol{ y} = \boldsymbol{k} * (\boldsymbol{\Lambda_H s}) + \boldsymbol{n}
\end{equation}
where the symbol * stands for linear convolution. In this article, the convolution is reversible as the kernel does not have zero points on discrete spectrum.

With the expression, the estimation of $\boldsymbol{s}$ becomes a non-blind deblurring problem. Further, we can regard the signal problem as an image deblurring problem, as the data plotted in time-sensor domain shows temporal and inter-sensor correlation like natural image, see Fig. \ref{fig:ZeroShot Comp}. 
A series of image deblurring solutions decompose the problem into deconvolution and denoise problem\cite{schuler_machine_2013,zhang_learning_2017,zhang_CNN_learning_2017}. 
They use local convolution, consider image prior and show efficiency and accuracy. Based on these methods, we propose our Wiener-denoiser network estimator in Figure \ref{fig:net}. 


\subsection{Deconvolution of Misaligned Samples}
A Wiener filter is used for deconvolution, followed by channel gain correction module in Eqn.\ref{equ:hcorr}. The output is initial estimation $\boldsymbol{\widetilde{s}}$:
\begin{equation}
    \widetilde{S_h}[k] = \frac{K^*[k]}{|K[k]|^2+\Phi_n[k]/\Phi_s[k]} Y[k] 
\end{equation}
\begin{equation}
    \boldsymbol{\widetilde{s}} =\boldsymbol{\Lambda_H^{-1}} \cdot \mathcal{F}^{-1}(\boldsymbol{S_h})_{[1:ML]}
    \label{equ:hcorr}
\end{equation}
Here $Y,K$ are fourier transform of $y, k $, $\mathcal{F}^{-1}$ is inverse fourier transform. Notice that kernel $K$ should be padded with zero to $M(L+1)-1$ length in time domain, to ensure circular convolution is equivalent to linear convolution. Signal-to-noise ratio (SNR) is used for approximation of $\Phi_n[k]/\Phi_s[k]$. 

\subsection{Deep Learning Based Denoiser}
Wiener filter still suffers from noise\cite{son_fast_2017} and it does not use data prior knowledge. In this case, we designed a deep learning denoiser to utilize the prior knowledge of sensor signal and to better estimate $\hat{s}_+$.

\paragraph{Network}
The most widely used deep learning network architectures for image deblurring and denoising are MLP\cite{schuler_machine_2013}, CNN\cite{zhang_CNN_learning_2017}, ResNet\cite{nimisha_blur-invariant_2017} and Encoder-Decoder(e.g., FCN\cite{zhang_learning_2017}, U-Net\cite{zhang2018dynamic}). In our application scenario, the network should be able to adapt to signals of different lengths and achieve high accuracy while ensuring real-time response. Meanwhile, the network should avoid complicated structures to prevent overfitting on limited datasets and simple signal patterns. Considering accuracy and efficiency we employ a simplified U-Net Network\cite{UNet_RonnebergerFB15} as denoiser. 

The input of U-Net $\boldsymbol{y}$, $\boldsymbol{\widetilde{s}}$ are reshaped into matrices: since temporal and inter-sensors correlation of sensor data can be better captured 2D matrix, see Fig. \ref{fig:ZeroShot Comp}. 
The network has only two downsample and two upsample branches for efficiency. Parameters of the network is 1,036K. 

\paragraph{Loss Function}
Our target is to estimate $\hat{s}_+$, the column sum of $\hat{s}$ matrix. We also want each symbol (pixel on $\boldsymbol{s}$) to be accurately estimated. Therefore, the network is trained by minimizing pixel-wise loss $\mathcal{L}_{pix}$ and target-wise loss $\mathcal{L}_{tar}$:

\begin{footnotesize}
\begin{equation}
\mathcal{L}=\mathcal{L}_{pix}+\mathcal{L}_{tar}=\frac{1}{M}\sum_{m=1}^{M}\sum_{l=1}^{L}\frac{1}{L}\left|s_{m}[l]-\hat{s}_{m}[l]\right|^{2}+\left|s_{+}[l]-\hat{s}_{+}[l]\right|^{2}
\end{equation}
\end{footnotesize}

\section{Experiments}
\subsection{Dataset and Training Settings}
The network is trained on a semiconductor gas sensors dataset, with 14 sensors in a sensor array for measurements of values of CO concentration, humidity and temperature\cite{burgues2018estimation}. The target is to evaluate a comprehensive gas metric with an average of local sensors' data, so we assume the OAC function is an average one. 

To simulate a sensor data OAC scenario, we randomly choose pieces of samples with $L=256$, $M=14$ in continues periods. For misaligned OAC channel settings, in training dataset we uniformly sampled $\tau_m$ within $T$, and $h_m$ from $e^{j0}$ to $e^{j\pi}$. The SNRs (use Es/N0 for symbols) are randomly chosen from -20dB to 20dB. In testing dataset, $\tau_m$, $h_m$ and SNRs are fixed since real world MAC channel does not change rapidly. The simulation outputs $\boldsymbol{y}$. We also calculated $s_+$ without misalignment as the ground truth in training. 6056 pairs of $s_+$ and $\boldsymbol{y}$ are collected for training and another 2153 pairs are used for testing. 

For network training, we use Adam optimizer, with learning rate 1$e$-4, batchsize $64$, epoches number 50. 

\subsection{MSE and Efficiency Comparison}
\begin{figure}[htbp]
  \centering
  \includegraphics[width=0.6\linewidth]{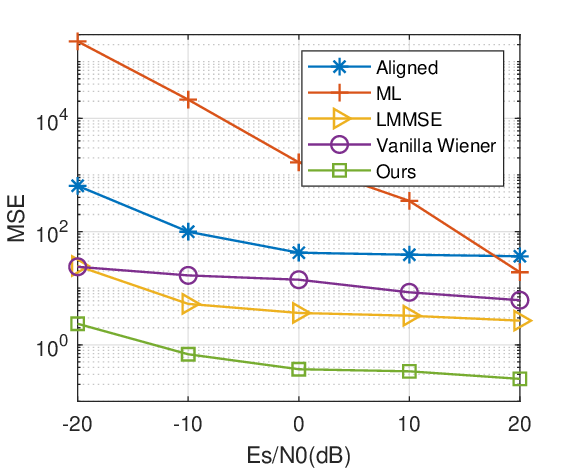}
  \caption{A Comparison of MSE under various SNR}
  \label{fig:MSE Comp}
\end{figure}

\begin{figure}[htbp]
  \centering
  \includegraphics[width=0.6\linewidth]{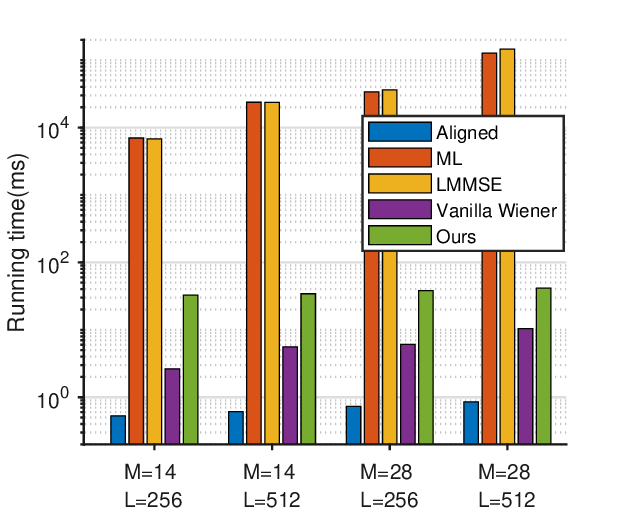}
  \caption{A Comparison of efficiency under various data scales}
  \label{fig:Eff Comp}
\end{figure}

We computed MSE to reveal whether estimators can accurate recover $s_+$ on the testing dataset. We also tested estimators on data with different number of sensors $M$ and packet length $L$, and focus on their efficiency in real time OAC. Here, we compared aligned estimation\cite{align_UAV}, ML estimation\cite{shao_federated_2022}, LMMSE estimation\cite{shao_bayesian_2021}, vanilla Wiener filter estimation\cite{wiener_extrapolation_1949} and our method, see Fig. \ref{fig:MSE Comp} and \ref{fig:Eff Comp}.  

A lower MSE indicates more accurate estimation. Our method outperforms other methods under different SNR settings. In low SNR regime, ML estimation performs worst, due to the error propagation effect that estimation error introduced by noise in a sample can propagate to other samples. LMMSE method performs slightly worse than our method, since it uses limited prior knowledge (mean and variance of each sensor data), it only promises that estimation obeys prior distribution but cannot avoid temporal disorder. The vanilla Wiener filter's result can be the ablation study of our denoiser network, our method obtains lower MSE than Wiener filter, indicating that our prior based denoiser network makes sense. 

Since channel-gain misalignment becomes dominate factor in the high SNR regime, estimators suffer from error floors that MSE drops slower. Similar phenomenon has also been reported in Shao et.al's work\cite{shao_bayesian_2021}. 

For efficiency, we recorded the mean running time of estimators. The aligned estimator runs fastest due to the simplest design. ML and LMMSE methods require large matrix computation, causing extremely long computation time. For our method and wiener filter, most computations are convolution on GPU. The running times of these methods are linear to the data scale. Based on data amount and processing time, our estimator throughput is about 358KB/s for sensors' data. 

In conclusion, our Wiener-Denoiser estimator shows advantage in accuracy and the efficiency is acceptable for real time computation, under different noise levels and data scales.

\subsection{Zero-Shot Performance of our Network}

\begin{figure}[h]
  \centering
  \includegraphics[width=0.8\linewidth]{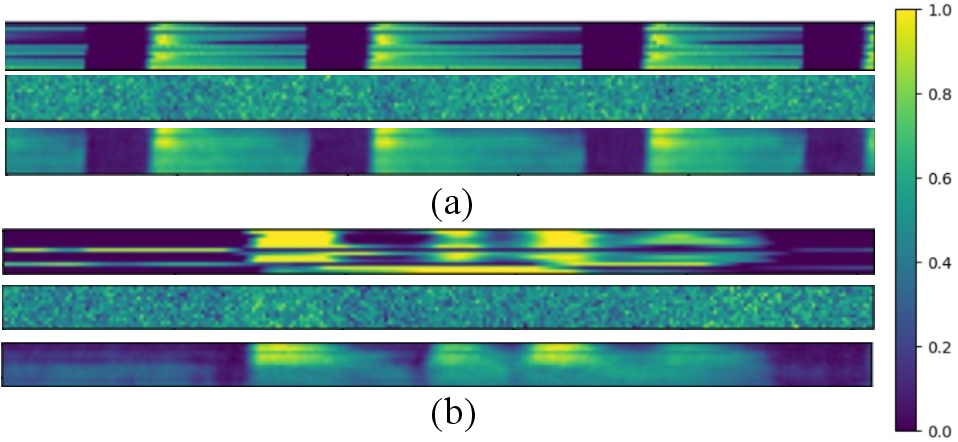}
  \caption{Visualize of the data. In each group, there are sensor data $\boldsymbol{s}$ (up), observed samples $\boldsymbol{y}$ at BS (middle) and our estimation $\hat{s}$ (low), the horizontal axis is time and the vertical axis is sensor id. (a) Semiconductor gas sensors data (b) Air quality sensors data}
  \label{fig:ZeroShot Comp}
\end{figure}

We also test our trained network on unseen sensor data to evaluate its zero-shot performance. Here we select a 13-sensor air quality data\cite{misc_air_quality_360}, shown in Fig \ref{fig:ZeroShot Comp}(b). 

\begin{table}[htbp]
  \centering
  \caption{A Comparison of MSE of estimation methods using different data (SNR=0dB)}
  \resizebox{\linewidth}{!}{
    \begin{tabular}{lccccc}
    \toprule
          & \multicolumn{1}{c}{Aligned } & \multicolumn{1}{c}{ML} & \multicolumn{1}{c}{LMMSE } & \multicolumn{1}{c}{Vanilla Wiener } & \multicolumn{1}{c}{Ours} \\
    \midrule
    Air quality & 31.93 & 1080.02 & 4.14  & 10.77 & \textbf{3.34} \\
    Original Data & 42.30  & 1666.23 & 3.64  & 14.09 & \textbf{0.36} \\
    \bottomrule
    \end{tabular}%
    }
  \label{tab:Zero-shot}%
\end{table}%

When SNR=0dB, the MSE on air quality data, original data are 3.34 and 0.36 separately, indicating that the network does not estimate as well as it performs on original dataset. The main problem is that the data features have changed, so the prior knowledge in network might help less. But compared with other methods in Table\ref{tab:Zero-shot}, it still achieves enhancement on MSE. We also measure the estimation SNR, if we see the estimation error as receiver noise. The estimation SNR are 9.9dB and 22.06dB on new and original data, acceptable for sensor data transmission. We reach a conclusion that our trained network can be used for unseen sensor data. 

\section{Conclusion}
In this paper we proposed a network to address channel-gain and time misalignment in OAC. We formulated the problem into an image deblurring problem, since the misalignment process is similar to image blurring. A U-Net denoiser is trained to exploit the prior knowledge from signal data for noisy deblurring task. Our method shows accuracy, efficiency, and zero-shot ability. Future works will focus on how to keep accuracy under different sources and sizes of data, and take into account of dynamic channel nature and inaccurate channel-gain estimation for improvement.

\bibliographystyle{ACM-Reference-Format}
\bibliography{sample-base}

\appendix

\end{document}